\documentclass[twocolumn,amsmath,amssymb,amsfonts,prb,floatfix, tightenlines,superscriptaddress]{revtex4}

\usepackage{graphicx} % Include figure files
\usepackage{dcolumn} % Align table columns on decimal point
\usepackage{bm} % bold math

\begin{document}

\title{Disorder-induced pseudodiffusive transport in graphene nanoribbons}  % Force line breaks with \\

\author{P.~Dietl}
\affiliation{Institut f\"{u}r Theoretische Festk\"{o}rperphysik, Universit\"{a}t Karlsruhe, D-76128 Karlsruhe, Germany}
\author{G.~Metalidis}
\affiliation{Institut f\"{u}r Theoretische Festk\"{o}rperphysik, Universit\"{a}t Karlsruhe, D-76128 Karlsruhe, Germany}
\affiliation {DFG Center for Functional Nanostructures (CFN), Universit\"at Karlsruhe, 76128 Karlsruhe, Germany}
\author{D.~Golubev}
\affiliation{Institut f\"{u}r Theoretische Festk\"{o}rperphysik, Universit\"{a}t Karlsruhe, D-76128 Karlsruhe, Germany}
\author{P.~San-Jose}
\affiliation{Department of Physics, Lancaster University, Lancaster, LA1 4YB United Kingdom}
\author{E.~Prada}
\affiliation{Department of Physics, Lancaster University, Lancaster, LA1 4YB United Kingdom}
\author{H.~Schomerus}
\affiliation{Department of Physics, Lancaster University, Lancaster, LA1 4YB United Kingdom}
\author{G.~Sch\"{o}n}
\affiliation{Institut f\"{u}r Theoretische Festk\"{o}rperphysik, Universit\"{a}t Karlsruhe, D-76128 Karlsruhe, Germany}
\affiliation {DFG Center for Functional Nanostructures (CFN), Universit\"at Karlsruhe, 76128 Karlsruhe, Germany}

\date{\today}% It is always \today, today,

\begin{abstract}
We study the transition from ballistic to diffusive and localized
transport in graphene nanoribbons in the presence of binary
disorder, which can be generated by chemical adsorbates or
substitutional doping. We show that the interplay between the
induced average doping (arising from the non-zero average of the
disorder) and impurity scattering modifies the traditional picture
of phase-coherent transport. Close to the Dirac point, intrinsic
evanescent modes produced by the impurities dominate transport at
short lengths, giving rise to a new regime analogous to
pseudodiffusive transport in clean graphene, but without the requirement of
heavily doped contacts. This intrinsic
pseudodiffusive regime precedes the traditional ballistic,
diffusive and localized regimes. The last two regimes exhibit a
strongly modified effective number of propagating modes, and a
mean free path which becomes anomalously large close to the Dirac
point.
\end{abstract}

\maketitle

\section{Introduction} 
Graphene continues to fascinate, due in particular to its
transport properties close to the Dirac point, where a finite
conductivity is observed experimentally~\cite{NovoselovNAT05} in
spite of the vanishing density of states. For clean graphene,
simple Dirac fermion models~\cite{GonzalezNPB93} prove to be
sufficient to describe this effect based on a gapless spectrum of
evanescent modes, pinned at the reservoir contacts, that allow for
quantum tunneling through macroscopically large graphene
samples.\cite{KatsnelsonEPJB06,TworzydloPRL06} Surprisingly, the
full transport statistics becomes indistinguishable from that of
diffusive metals,\cite{TworzydloPRL06} for which reason  the name
`pseudodiffusive' was coined to describe this transport regime. Furthermore,
these evanescent modes can give rise to Hanbury Brown-Twiss type
cross-correlations in a ballistic multiprobe graphene structure,\cite{Laakso2008}
similar to those observed in conventional diffusive conductors.

Just as ballistic transport, the effects of disorder in graphene display a
surprising richness. E.g., armchair ribbons are found to be more sensitive to
bulk disorder as their zigzag counterparts, while edge disorder can open up
transport gaps in both of them.\cite{Mucciolo2009} In general, different types of
disorder can be classified in terms of preserved symmetries, including chirality
which is preserved for disorder that is smooth on the scale of interatomic
distances \cite{McCannPRL06, AleinerPRL06,OstrovskyPRB06, SchuesslerPRB09} and
significantly modifies the standard results for transport in disordered metals.
For example, carriers in graphene ribbons with smooth disorder cannot be
localized,\cite{NomuraPRL07} and their conductivity is seen to increase with
system size $L$ as $\sigma \sim \sqrt{L/l_m}$ in 1D,\cite{Titov2007} and $\sim
\ln{L/l_m}$ in 2D, where $l_m$ is the mean free
path.\cite{SanJosePRB07,BardarsonPRL07} Short-range disorder, on the other hand,
does not preserve chirality, and is therefore generally understood to induce a
sequence of ballistic transport, diffusion, and Anderson localization in quasi-1D
wires of increasing length.\cite{OstrovskyPRB06,CrestiNR08, AreshkinNL07}

%Furthermore, the ballistic regime is understood to
%be non-universal, while diffusion and localization display
%universality and are conveniently described by  random matrix
%theory and the Dorokhov-Mello-Pereyra-Kumar (DMPK) %equation.\cite{DorokhovJETP82,MelloAP88,BeenakkerRMP97} %However, within
%these approaches transport through evanescent modes is %neglected.

Until now, the influence of evanescent modes close to the Dirac point has been studied in ballistic systems only, and it is therefore unclear whether their peculiar spectrum could manifest itself in
some significant way in the conductivity of disordered graphene
nanoribbons.
\begin{figure}
\includegraphics[width = 8cm]{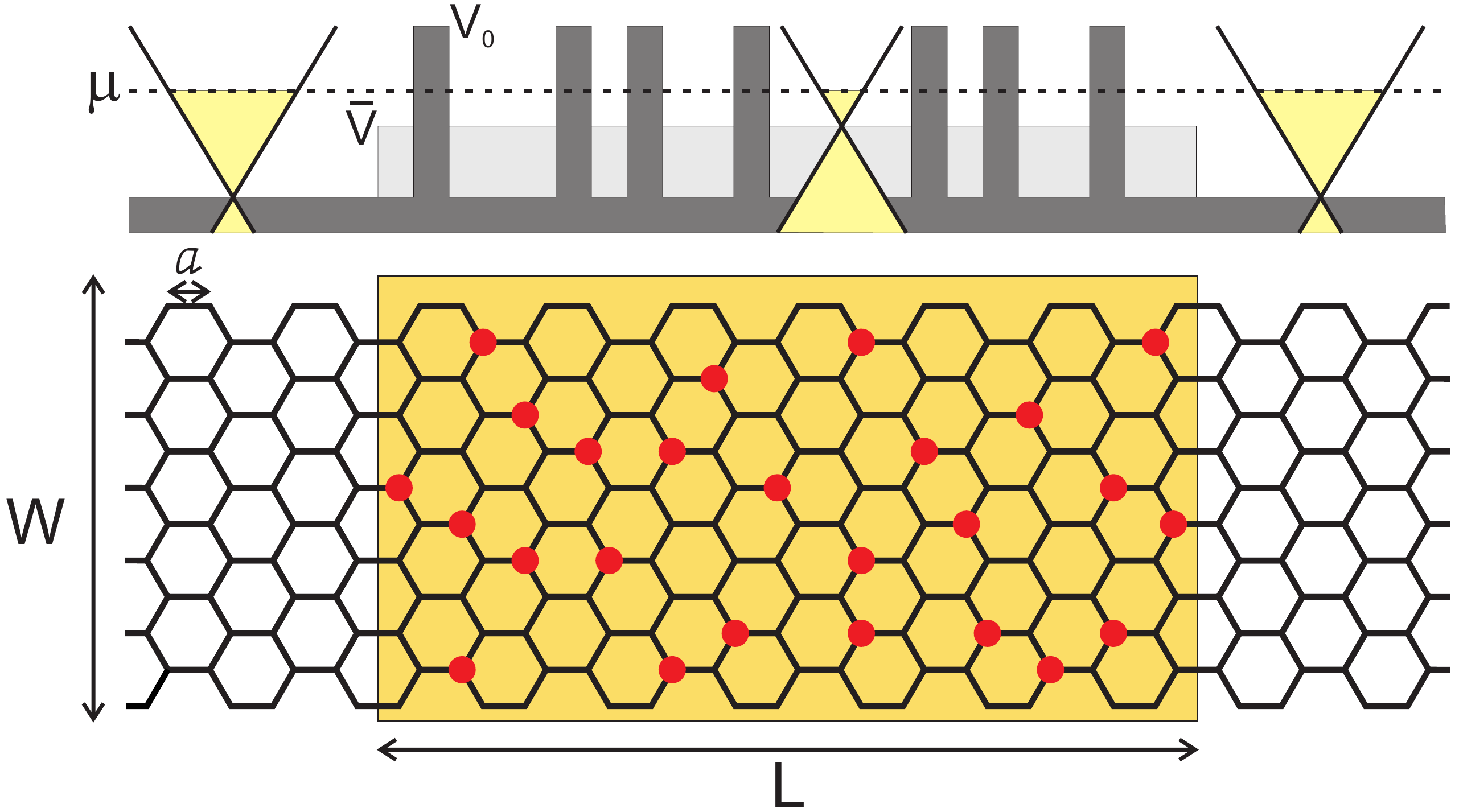}
\caption{(Color online) Schematic view of an armchair ribbon with width $W$.
Disorder is introduced in a region of length $L$ by adding a
potential $V_0$ to a percentage $p$ of lattice sites. Disorder
sites are depicted by red dots. The induced local doping, due to
the non-zero average of the disorder, is effectively described by
a barrier of height $\bar{V}=p V_0$.\label{figsetup}}
\end{figure}
In this work we will show that evanescent modes play a crucial role for the
transport in graphene nanoribbons with short-range disorder. In particular,
intrinsic evanescent modes arising from the disorder itself (i.e., independently of the
contacts) can reinstate an `intrinsic pseudodiffusive' regime which {\em
precedes} the ballistic, diffusive and localized regimes. This new regime arises
if the short-range disorder does not vanish on average and therefore induces
systematic doping, as is the case, e.g., for chemical adsorbates
\cite{RobinsonPRL08} or substitutional doping.\cite{LherbierPRL08} We describe
such systems using a binary disorder model (the random binary alloy
\cite{PantJPC80, DunlapPRL90}) and obtain our results by combining numerical
computations with analytical considerations. For longer ribbons, the intrinsic
pseudodiffusive regime crosses over into the traditional diffusive and localized
regimes, but we find that the mean free path becomes anomalously large at an
effective Dirac energy which takes the induced doping into account.

%We further analyze
%the crossover from the intrinsic pseudodiffusive to the
%traditional diffusive and localized regimes and find that the %mean
%free path becomes anomalously large at an effective Dirac %energy
%which takes the induced doping into account.

\section{Binary Disorder Model} \label{sec:model}
We base our investigation of binary disorder in a graphene
nanoribbon on the following tight-binding Hamiltonian,
\begin{equation}
H = -t \sum_{\langle i,j\rangle} c_i^\dag c_j + \sum_{i} v_i
c_i^\dag c_i.
\end{equation}
The first term  describes clean graphene with hopping amplitude
$t\approx 2.9 eV$ between nearest neighbors on the hexagonal
lattice of carbon atoms (with interatomic distance $a=1.49$ \AA).
The second term describes binary disorder, generated by onsite
energies $v_i$ which are uncorrelated random variables that take
values $v_\textrm{high}=V_0$ with probability $p$, and
$v_\textrm{low}=0$ with probability $1-p$. We have dropped spin
indices in the model since spin is a passive degree of freedom in
all our discussion.

Physically, this model describes, for example, a graphene
nanoribbon with a fraction $p$ of its sites coupled to identical
chemical species adsorbed on its surface. If these adsorbates are
screened their presence is equivalent to a (real and energy
dependent) self-energy term acting on each of the carbon sites
coupled to an adsorbate.\cite{RobinsonPRL08} Typical values of
$V_0$ can be estimated from DFT calculations,\cite{GiovannettiPRL08} and for species such as Ag, Cu or Au,
one finds values $V_0 \approx 0.5 \text{eV} \approx 0.2 t$ which
are small  as compared to $t$.

\section{Results and discussion}
To probe the intrinsic transport properties in the presence of
binary disorder we present results of numerical computations on
nanoribbons of width $W$ and a disordered region of length $L$,
with the reservoir contacts placed at $x=\pm \infty$ (see Fig.
\ref{figsetup}).  These computations are based on the recursive
Green's function technique.\cite{MetalidisPRB05,SanvitoPRB99,RobinsonPRB07} We obtain qualitatively similar results for metallic and
semiconducting armchair ribbons as well as for zigzag ribbons; we
therefore only show results for metallic armchair ribbons.

\begin{figure}
\includegraphics[width = 8cm]{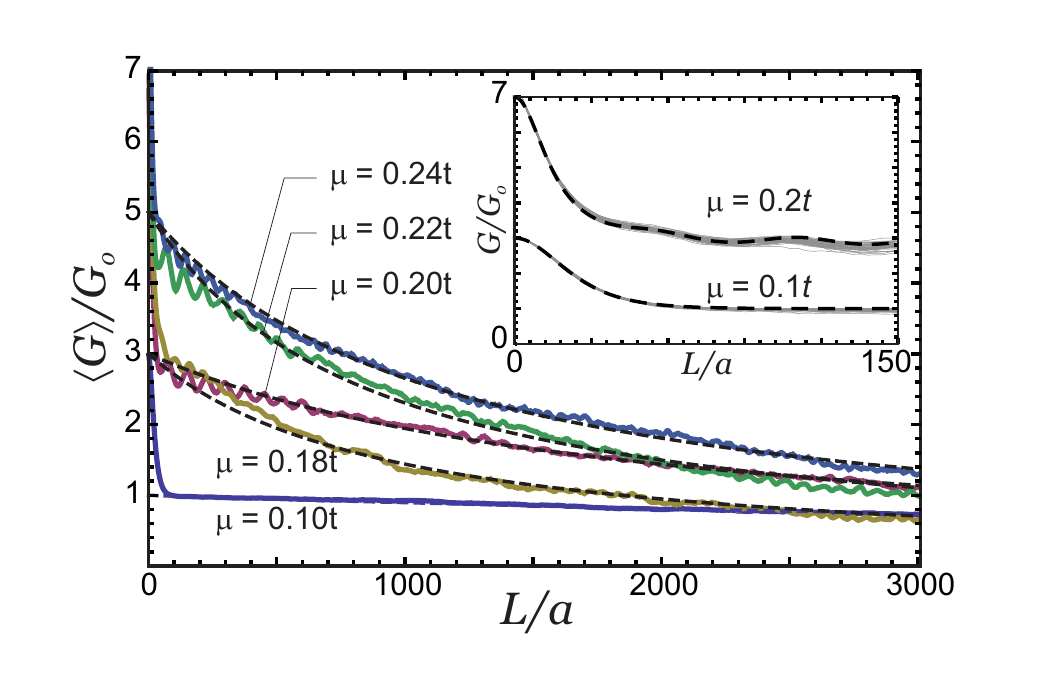}
\caption{(Color online) Averaged conductance $\langle G \rangle$ (in units $G_0 =
\frac{2e^2}{h}$) of an armchair ribbon of width $W = 103 \sqrt{3}
a/2$ as a function of the length of the disordered region for different chemical potentials. The
parameters of the binary disorder are $p=0.4$, $V_0=0.25t$, so
that the effective Dirac point in the disordered region is shifted
from zero to $\bar{V}=pV_0=0.1 t$. Dashed curves are fits to Eq. (\ref{Gballistic}), with $l_m$ as a fit parameter. Inset:
Pseudodiffusive regime zoom-in. Thin grey curves correspond to
different disorder realizations. Good agreement is obtained with
the ballistic conductance through a square potential barrier of
height $\bar{V}$, shown in dashed black, see Eq. (\ref{Gbarrier}).
\label{fig:GvsL}}
\end{figure}

Figure \ref{fig:GvsL} shows the averaged conductance of a ribbon of
width $W =103 \sqrt{3} a/2$ as a function of the length of the
disordered region; different curves correspond to different values
of the chemical potential. The disorder parameters used, $p=0.4$
and $V_0=0.25 t$, correspond to a substantial concentration of
adsorbates that are weakly hybridized with their carbon hosts; a
sample size of $50$ impurity configurations is used in the
average.

%\begin{figure}
%\includegraphics[width = 8cm]{GvsL2.pdf}
%\caption{Conductance versus length of a metallic armchair nanoribbon of width $W=103\sqrt{3}a/2$ for two values of
%the reservoir chemical potential. Disorder parameters are as in Fig. \ref{figGvsL}. Thin grey lines correspond to
%different disorder realizations, and thick red correspond to a disorder average. Good agreement is obtained with
%the ballistic conductance through a square potential barrier of height $\bar{V}$, shown in dashed black, see Eq. (\ref{Gbarrier}).
%\label{fig:GvsL2}}
%\end{figure}

\begin{figure}
\includegraphics[width = 8cm]{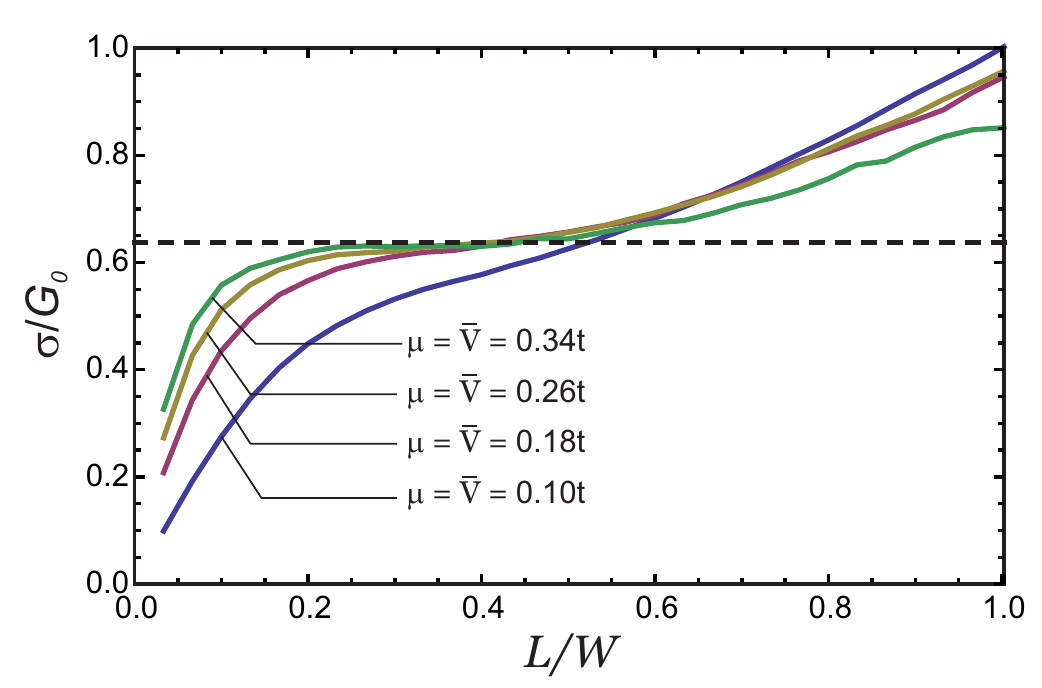}
\caption{(Color online) Conductivity  $\sigma = \langle G\rangle L/W$ versus length for an
armchair ribbon of width $W = 103 \sqrt{3} a/2$ at the effective
Dirac point $\mu=\bar{V}$ and fixed impurity fraction $p=0.4$.
Different curves correspond to different strength $V_0=\bar V/p$
of the disorder. As the chemical potential of the effective Dirac
point increases, the number $N$ of incoming modes grows, and a
plateau develops at the universal conductivity minimum $2G_0/\pi$
(dashed black line). \label{figsigma}}
\end{figure}

Two distinct transport regimes immediately catch the eye. The
conductance first decreases sharply over a length scale comparable
to the ribbon width $W$, and then abruptly crosses over to a much
slower decay. The conductivity $\sigma=\langle G\rangle L/W$ in the pseudodiffusive regime, shown in Fig. \ref{figsigma}, develops a clear
plateau at the value $\sigma=2G_0/\pi$ (the minimal conductivity
in  clean graphene \cite{KatsnelsonEPJB06,TworzydloPRL06}) when
the chemical potential $\mu$ coincides with the average value
$\bar V=p V_0$ of the disorder. We now argue that this transport
regime indeed constitutes an {\em intrinsic} pseudodiffusive
regime arising from disorder-induced evanescent modes, while the
crossover to the truly diffusive regime is characterized by a mean
free path which becomes anomalously large at $\mu=\bar V$.

\subsection{Intrinsic pseudodiffusive regime}
To understand the numerical results for short lengths it is useful
to compare the average conductance of Fig. \ref{fig:GvsL} to the
conductance of the same nanoribbon in which the disordered region
is replaced by a potential barrier of height equal to the average
disorder doping $\bar{V}=p V_0$. We compute the conductance
through the barrier, $G_b$, by wave-matching within the Dirac
equation description of low-energy transport, which yields
\begin{equation} \label{Gbarrier}
    G_b=G_0\sum_n\left|\frac{\left(1-z_{nk}\right)^2\left(1-z_{nk'}\right)^2}{e^{ik'L}\left(z_{nk}-z_{nk'}\right)^2+e^{-ik'L}\left(1-z_{nk}z_{nk'}\right)^2}\right|^2.
\end{equation}
Here the conductance quantum is defined by $G_0=2e^2/h$, the sum
runs from $n=-(N-1)/2$ to $(N-1)/2$, the number of incoming
propagating modes is $N=1+2\textrm{Int}[\mu W/(\pi\hbar v)]$,
$\mu$ is the reservoir chemical potential, the Fermi velocity is
given by $v=\frac{3}{2} t a/\hbar$,
$z_{nk}=(k+iq_n)/\sqrt{k^2+q_n^2}$, $k=\sqrt{\left(\mu/\hbar
v\right)^2-q_n^2}$, $k'=\sqrt{\left[(\mu-\bar{V})/\hbar
v\right]^2-q_n^2}$, and $q_n=n\pi/W$.

The comparison between $G$ and $G_b$ over the range of the abrupt
decay is shown in the inset of Fig. \ref{fig:GvsL}, where each
grey curve corresponds to an individual disorder configuration, and the dashed black curve corresponds to Eq. (\ref{Gbarrier}). We
obtain a remarkably good agreement of both curves, even down to
hallmark features such as Fabry-P\'erot oscillations. This good
agreement only begins to break down at lengths for which the
disorder starts to mix modes, i.e., the mean free path $l_m$,
which will be discussed below.

The implications are clear: The sharp decline of conductance observed at lengths
$L\lesssim W$ is due to the decay of evanescent modes induced by the sudden
change in local average potential $\bar{V}$. This average potential shift is well
defined because the effect of the disorder is self-averaging at the considered
impurity concentrations.. Although these evanescent waves are produced by the
impurities, their combined effect is equivalent to evanescent modes pinned at the
boundary of the disorder region, $x=0$ and $x=L$, which allow tunneling through
an effective potential barrier $\bar{V}=p V_0$. Furthermore, their contribution
to transport at energies $\mu\sim\bar{V}$ and very wide ribbons produces the
universal statistics of diffusive metals. Therefore, the observed fast decay in
average conductance is the signature of a disorder-induced pseudodiffusive
regime. As we will show in the following section, the selfaveraging of the
disorder arises since the mean free path becomes anomalously large around the
effective Dirac point $\mu=\bar{V}$. Contrary to the ballistic regime, the
intrinsic pseudodiffusive regime is therefore universal in the limit of wide and
short ribbons, which is verified by the negligible spread of the curves in the
inset of Fig. \ref{fig:GvsL} over the range of the abrupt decay.

\subsection{Ballistic to diffusive crossover}
At lengths $L\sim W$ evanescent modes through the disordered
region have decayed, and give way to ballistic propagation of the
remaining $N_p(\mu)=1+2\textrm{Int}[(\mu-\bar{V})W/(\pi\hbar v)]$
modes, up to lengths $L$ comparable to the mean free path $l_m$,
where the traditional diffusive regime sets in. In the range
$W<L<N_pl_m$, therefore, the large-$N_p$ form of the average
conductance in the ballistic to diffusive crossover becomes
approximately valid,\cite{BeenakkerRMP97}
\begin{equation} \label{Gballistic}
   \langle G\rangle=G_0\frac{N_p}{1+L/l_m}.
\end{equation}
This is followed by a transition into a localized regime for nanoribbon lengths greater that the localization length $\xi=N_pl_m$,
for which $\langle G\rangle<G_0$. From that point on, $\langle G\rangle$ decays exponentially as $\exp{(-2L/\xi)}$.

\begin{figure}
\includegraphics[width = 8cm]{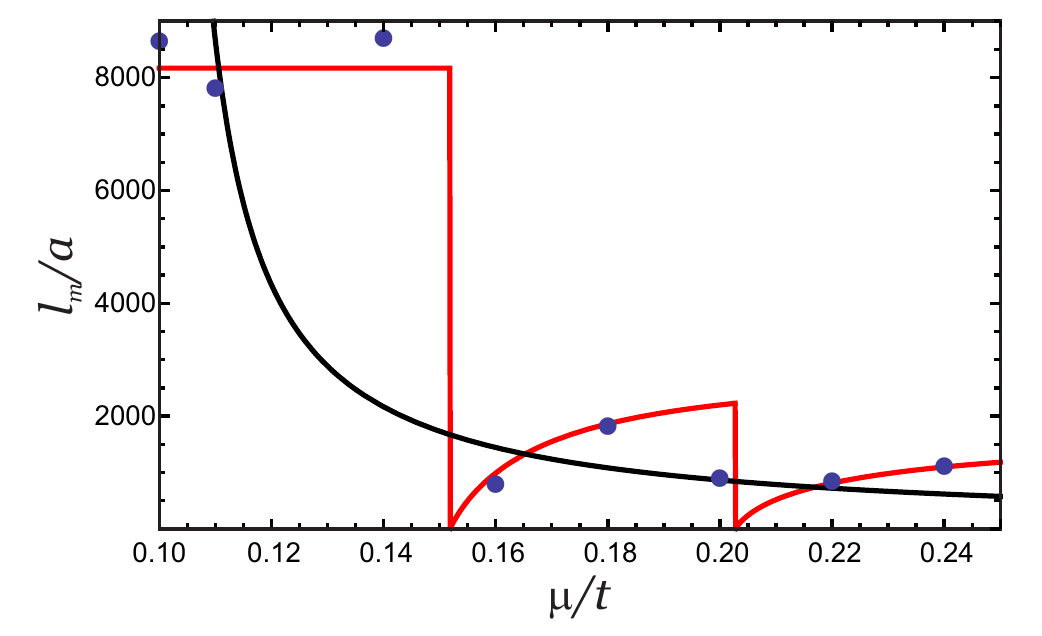}
\caption{(Color online) Mean free  path as a function of chemical potential for a
metallic armchair nanoribbon of width $W=103\sqrt{3}a/2$, impurity
strength $V_0=0.25 t$, and impurity fraction $p=0.4$. Black curve:
bulk value, Eq.\ (\ref{eqmeanfreepath}). Red curve: finite width
result, Eq.\ (\ref{lm_cresti}). Dots: values obtained by fitting
numerical results to Eq.\ (\ref{Gballistic}).\label{fig:lmfit}}
\end{figure}

The mean free path in the bulk graphene sample is known to be inversely proportional
to the electron energy and the disorder strength, see e.g. Refs. \onlinecite{ShonJPSJ98,SuzuuraPRL02}.
Adapting the results of these
references to our binary disorder model, we arrive at the following expression for the mean free path
\begin{equation} \label{eqmeanfreepath}
l_m = \frac{3 \sqrt{3}}{4} \frac{a t^3}{p (1-p) V_0^2 |\mu-\bar{V}|}.
\end{equation}
The $1/|\mu-\bar{V}|$ divergence can be traced back to the
vanishing bulk density of states of graphene at $\mu\approx
\bar{V}$ in the presence of average doping $\bar{V}=pV_0$. This
dependence is consistent with the behavior observed in Fig.
\ref{fig:GvsL} for lengths greater than $W$. However, a detailed
analysis shows that Eq. (\ref{eqmeanfreepath}) provides only a
qualitative description of our data. In order to improve the
agreement one has to account for the finite size effects in the
density of states. In the simplest weak
disorder approximation the mean free path acquires the
form~\cite{CrestiNR08}
\begin{eqnarray}\label{lm_cresti}
l_m = \frac{\pi \sqrt{3}}{4} \frac{t^3W}{p (1-p) V_0^2
}\frac{N_p}{\rho^2(\mu-\bar{V})},
\end{eqnarray}
where the effective density of states $\rho(\mu)$ is defined as
follows
\begin{eqnarray}
\rho(\mu)&=&
1+2|\mu|\sum_{n=1}\frac{\theta(\mu^2-E_n^2)}{\sqrt{\mu^2-E_n^2}}.
\end{eqnarray}
Here $E_n= \hbar v n\pi/W$ are threshold
energies at which new channels open up. According to Eq.
(\ref{lm_cresti}) the mean free path is no longer divergent at
$\mu=\bar{V}$, rather it saturates at a finite value
$l_{\max}\propto W$. In Fig.\ \ref{fig:lmfit} we show the
comparison of the mean free path Eq.\ (\ref{lm_cresti}) to the one
resulting from fitting our data in Fig.\ \ref{fig:GvsL} with Eq.
(\ref{Gballistic}); we find that Eq.\ (\ref{lm_cresti}) reproduces
our numerical data very accurately.

%Using Eqs. (\ref{Gballistic}) and (\ref{eqmeanfreepath}) we obtain a good agreement with the numerical
%conductance in this regime, see dashed lines in Fig. \ref{figGvsL}. Note, however, that although for $\mu=\bar{V}$, Eq. (\ref{eqmeanfreepath})
%predicts an $L$-independent average conductance (since $l_m\rightarrow \infty$), we do observe a very slow decay with increasing length
%at this energy (not shown). Indeed, for finite width nanoribbons it should be expected that the mean free path divergence is cut off,
%while remaining anomalously large at the Dirac point~\cite{CrestiNR08}. Note that only metallic nanoribbons will exhibit such a long
%lived conducting channel at $\mu=\bar{V}$, while semiconducting ones that lack modes at zero lateral momentum will have a vanishing
%conductance at this point.

\subsection{Dependence of conductance on disorder parameters}
\begin{figure}
\includegraphics[width = 8.5cm]{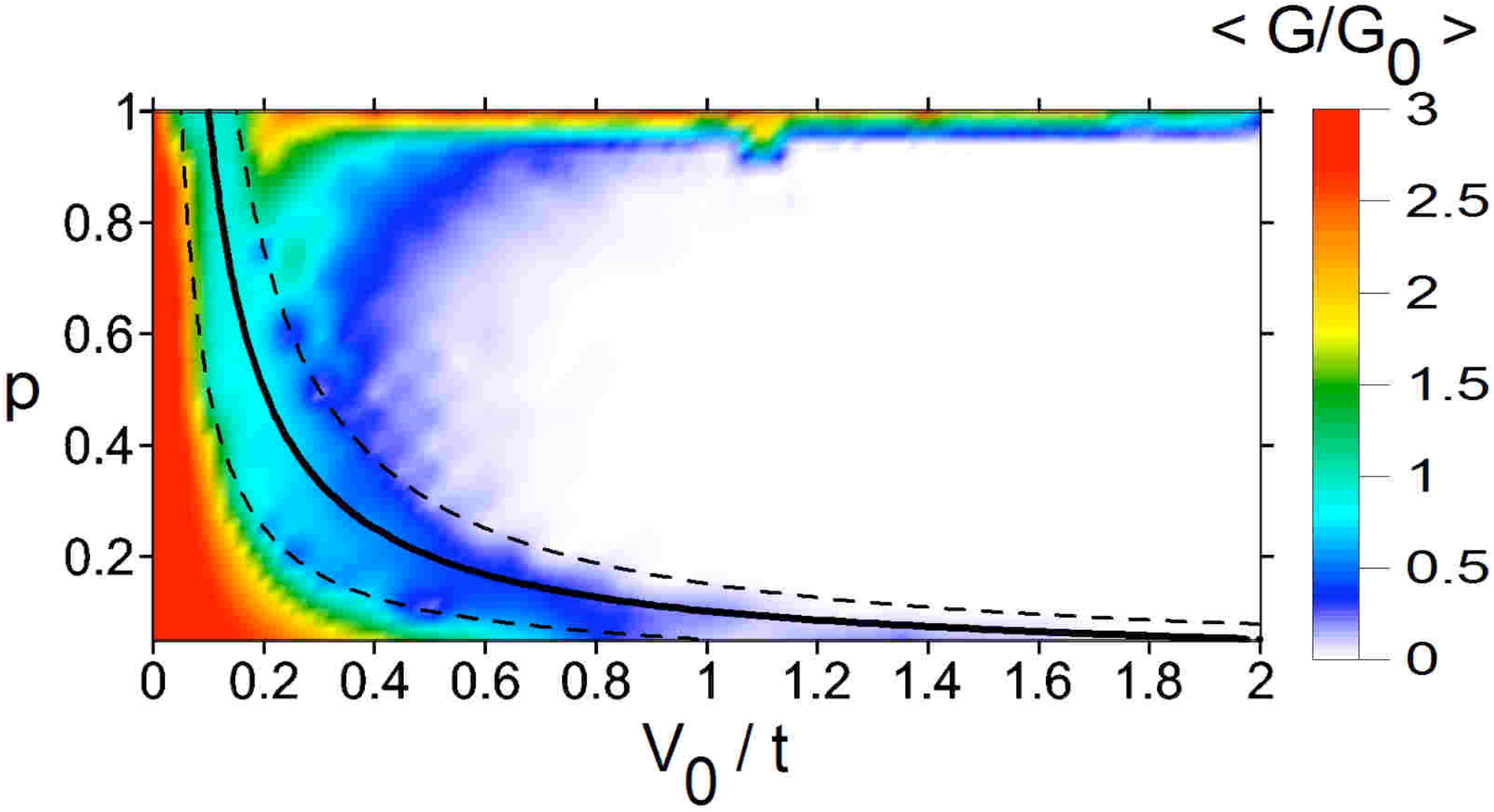}
\caption{(Color online) Conductance of a graphene armchair ribbon of length
$L=3000 a$ and width $W=103 \sqrt{3} a/2$ for different values of
the impurity fraction $p$ and scattering potential $V_0$. The
chemical potential is fixed to $\mu = 0.1 t$. The solid line corresponds
to the condition  $\mu=\bar{V}=p V_0$. The dotted lines are guides for the eye, bounding the
region of increased conductance around the effective Dirac energy.
\label{figphasediagram}}
\end{figure}
In order to give a comprehensive overview of the interplay between
the various disorder parameters, we have calculated the
conductance of a long metallic graphene ribbon ($L=3000 a$) for
different parameters $p$ and $V_0$, at a fixed chemical potential
$\mu=0.1 t$. The results are summarized in
Fig.~\ref{figphasediagram}. Parameters $p$ and $V_0$ that lead to
a large mean free path show up as regions with a relatively high
conductance. E.g., in the region around $p V_0 = \mu$ depicted
with a solid black curve in Fig.~\ref{figphasediagram}, the
conductance is close to $G = 2e^2/h$, even for this rather long
ribbon. The conductance is also large around $p=0$ and $p=1$,
which correspond to the regimes where the effective disorder
strength $p(1-p)V_0^2$ is small. This factor thus accounts for the
semilunar shape of the phase diagram. Nevertheless, the physical
picture leading to a high conductance is somewhat different for
small and high disorder percentages. For small $p$ (and $V_0$), we
are in the strict ballistic regime, where the electrons do not
feel the impurities at all, not even their average doping effect.
On the other hand, for high $p$, we have effectively a
disorder-induced potential barrier in the system.

%As long as this barrier is smaller than the Fermi energy ($V_0 < \mu$), electrons propagate freely over the barrier and give a high conductance. If the barrier becomes higher than the Fermi energy ($V_0 > E_F$), one would expect the conductance to decrease fast. However, due to Klein tunneling, electrons will not be scattered, but they can just propagate straight through the barrier by occupying a hole state. (Klein tunneling in graphene is studied in detail in Ref.~\onlinecite{Katsnelson2006}). As the percentage of scatterers is decreased a bit, the short-range character of the single impurities becomes apparent, and Klein tunneling quickly vanishes, leading to a zero conductance.

\section{Conclusion}

In conclusion, we considered the conductance of graphene nanoribbons containing
binary disorder (typical for chemical doping) which induces a
shift $\bar{V}$ of the effective Dirac point, along with
scattering on single discrete impurities. For short ribbons we
identified a transport regime dominated by evanescent modes which
give rise to an intrinsic form of pseudodiffusive transport. At a
longer length scale, a ballistic, and then a diffusive regime is
entered where the mean free path is determined both by scattering
on the individual impurities, as well as by the shifted potential.
Close to the effective Dirac point, the mean free path becomes
anomalously large, which ensures the universality of the intrinsic
pseudodiffusive regime. Compared to ordinary disordered
conductors, the presence of non-zero-average short-range disorder
typical for chemically doped graphene therefore adds a new stage
to the conventional path towards localization. The predicted intrinsic pseudodiffusion should be directly measurable in chemically functionalized graphene nanoribbons.

This research was supported by the European
Commission via Marie Curie Excellence Grant MEXT-CT-2005-02377.

\bibliographystyle{apsrev}
\bibliography{paper}

\end{document}